\documentclass[runningheads]{llncs}
\usepackage{graphicx}
\usepackage{booktabs}
\usepackage{cite}
\usepackage{amsmath,amssymb,amsfonts}
\usepackage{algorithmic}
\usepackage{textcomp}
\usepackage{xcolor}
\usepackage{boo ktabs}
\usepackage{array}\newcolumntype{P}[1]{>{\centering\arraybackslash}p{#1}}

\usepackage{lipsum,lineno}

\begin{document}

\title{Exploring a Handwriting Programming Language for Educational Robots\thanks{Supported by the NCCR Robotics, Switzerland}}

\author{Laila El-Hamamsy\inst{1,2}\orcidID{0000-0002-6046-4822} \and
Vaios Papaspyros\inst{1}\orcidID{0000-0001-8257-1891} \and
Taavet Kangur\inst{1}
\and
Laura Mathex\inst{1}
\and
Christian Giang\inst{3,5}\orcidID{0000-0003-2034-9253}
\and
Melissa Skweres\inst{2}
\and
Barbara Bruno\inst{4}\orcidID{0000-0003-0953-7173}
\and
Francesco Mondada\inst{1,2}\orcidID{0000-0001-8641-8704}
}

\authorrunning{L. El-Hamamsy et al.}

\institute{MOBOTS Group, Ecole Polytechnique Fédérale de Lausanne (EPFL), Switzerland \and
Center LEARN, Ecole Polytechnique Fédérale de Lausanne (EPFL), Switzerland \and
D-VET Laboratory, Ecole Polytechnique Fédérale de Lausanne (EPFL), Switzerland \and
CHILI Laboratory, Ecole Polytechnique Fédérale de Lausanne (EPFL), Switzerland \and
SUPSI-DFA, Locarno, Switzerland
\\
\email{firstname.lastname@epfl.ch}}

\maketitle
\begin{abstract}
Recently, introducing computer science and educational robots in compulsory education has received increasing attention. However, the use of screens in classrooms is often met with resistance, especially in primary school. To address this issue, this study presents the development of a handwriting-based programming language for educational robots. Aiming to align better with existing classroom practices, it allows students to program a robot by drawing symbols with ordinary pens and paper. Regular smartphones are leveraged to process the hand-drawn instructions using computer vision and machine learning algorithms, and send the commands to the robot for execution. To align with the local computer science curriculum, an appropriate playground and scaffolded learning tasks were designed. The system was evaluated in a preliminary test with eight teachers, developers and educational researchers. While the participants pointed out that some technical aspects could be improved, they also acknowledged the potential of the approach to make computer science education in primary school more accessible. 

\ifdefined\USEIEEE
\else 
\fi 
\end{abstract}

\section{Introduction}

With the increasing efforts to introduce computer science (CS) in K-12 worldwide \cite{el-hamamsy_computer_2020}, it has become necessary to develop tools and environments that offer an appropriate progression across schools years \cite{papavlasopoulou_reviewing_2017}. Nowadays, aside from \textit{Text-based Programming Languages} such as Python or C, three main approaches to teach CS exist: \textit{CS unplugged activities}\footnote{CS Unplugged activities: \url{csunplugged.org/en/}} that do not involve the use of screens and aim at teaching core CS concepts rather than programming per se; \textit{Tangible Programming Languages} (TPLs) which involve the use of physical manipulatives to program virtual or physical agents \cite{melcer_bots_2018, mussati_tangible_2019}; and \textit{Visual Programming Languages} (VPLs) such as Scratch\footnote{See \url{scratch.mit.edu} for Scratch resources and \url{scratchjr.org} for Scratch Jr}, which offer block-based solutions to programming, either on tablets or computers \cite{horn_comparing_2009}. 
Although VPLs are increasingly present in education settings, they are met with resistance from teachers at the lower levels of primary school mainly due to the presence of screens \cite{el-hamamsy_computer_2020}. 
Conversely, TPLs, which are still not present in formal settings \cite{mehrotra_introducing_2020} appear to be better suited for the needs of primary school education. Indeed, the tangibility of TPL approaches aligns nicely with early childhood education practices and principles \cite{elkin_programming_2016, sapounidis_educational_2017}, promotes active and collaborative learning \cite{melcer_bots_2018, sapounidis_tangible_2019}, and is less likely to be met with resistance by teachers.
Moreover, TPLs seem to appeal more than VPLs to young students \cite{horn_comparing_2009} and help mitigate gender stereotypes \cite{horn_comparing_2009, melcer_bots_2018}. Lastly, experiments suggest that TPLs have a positive impact on interest, enjoyment and learning \cite{melcer_bots_2018}, all the while reducing the amount of errors produced \cite{sapounidis_evaluating_2015} and the amount of adult support required \cite{mehrotra_introducing_2020} compared to VPLs.

Educational Robots (ER), which by design embody and enact abstract CS concepts and directives, can be thought of as naturally aligned with the principles of TPL, and thus a potential means to promote the adoption of TPL by teachers. 
Among the solutions proposed in the literature to combine ER and TPL, those not requiring the purchase of additional dedicated hardware (i.e., beside the robot itself) seem particularly promising for primary school education settings \cite{horn_designing_2007}. A recent example is the work of Mehrotra et al. \cite{mehrotra_introducing_2020}, who developed a \textit{Paper-based Programming Language} (PaPL) for the Thymio II robot\footnote{Thymio II - \url{https://www.thymio.org/}}, observing that the paper-based interface helps promote the use of tangible interfaces in classrooms by offering a ubiquitous and accessible solution which only requires the robot and a classroom laptop. 
Similarly, TPL approaches that introduce handwriting as a way to provide instructions/values \cite{sabuncuoglu_kart-_2020} seem particularly interesting for primary school education settings, enhancing the programming sessions with the possibility to train essential graphomotor skills \cite{ziviani_development_2006}, which have been shown to be linked to improved student learning \cite{10.3389/fpsyg.2020.01810}.
 
In this work, we build on the principles of the first handwriting TPL approaches and the Paper-based Programming Language to introduce a \textit{Handwriting-based Programming Language} (HPL). By allowing students to program the robot by handwriting instructions on a sheet of paper, HPL relies on two cornerstones of primary school classrooms: the use of paper as material and of handwriting as a means of expression. HPL can thus be seen as an effort to take programming one step closer to primary school practices, all the while capitalising on the benefits handwriting has with respect to student learning \cite{10.3389/fpsyg.2020.01810}.

\section{Development of HPL} 
\label{sec:HPL_developemnt}

\begin{figure}
    \centering
    \vspace{-10pt}
    \includegraphics[width=\linewidth]{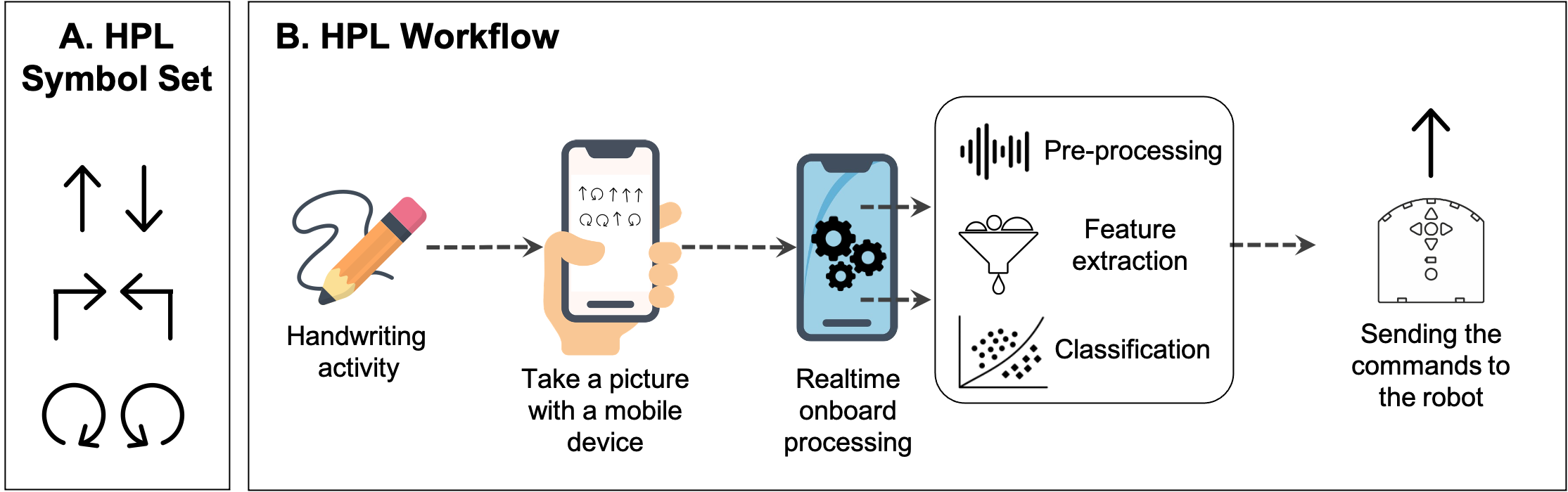}
    \vspace{-10pt}
    \caption{HPL symbols (A) and workflow from the handwriting activity to the robot commands (B). The ``up'' and ``down'' arrows (first row) correspond to forwards and backwards motions. The ``forward right'' and ``forward left'' arrows (second row) represent a forward motion along an arc of given circumference. The ``rotate right'' and ``rotate left'' arrows (third row) correspond to a rotation. Icons are taken from FlatIcon and the NounProject.}
    \vspace{-15pt}
    \label{fig:HPL_workflow}
\end{figure}

Our goal is to develop a Handwriting-based Programming Language for the Thymio II robot which could serve as a precursor to more advanced TPLs and VPLs for ER. For this reason, we focus on sequential programming and basic robot movements.  
As movements in TPLs are often represented by arrows, the instructions of the HPL were designed to allow for a ``clear mapping between [the] tangible and virtual commands'' \cite{papavlasopoulou_reviewing_2017} (see Fig. \ref{fig:HPL_workflow} - A). 

A data set of 6888 images of handwritten arrows was acquired from 92 students in the 9 to 12 age range, and split 60\% - 40\% between training and testing. 
Both during training and testing, each image is first pre-processed using adaptive filtering and binarisation followed by morphological operators, to reduce the noise while preserving edges and remove asperities \cite{herrera-camara_flow2code_2017}. 
The arrows are then isolated by identifying contours. The feature set is created to characterise the shape differences between the arrows by combining a) Fourier descriptors, b) Hellinger distances between the pixel density histograms along the x and y axes of the arrow to be classified and the centroid of the classes computed on the training set, and c) geometric features (circularity, convexity, inertia, rotated bounded rectangle, minimum including circle, moments, etc...). 
A number of state-of-the-art classifiers were trained and tested, namely: Decision Tree; Support Vector Machine; Multilayer Perceptron (MLP); kNN; Random Forest; AdaBoost; Naive Bayes classifier. The MLP achieved the best performance (reported in Table \ref{tab:conf_mat}).

\begin{table}
    \centering
    \vspace{-5pt}
    \caption{Confusion Matrix, Precision (P), Recall (R) and F$_1$-Score (F$_1$) for the MLP classifier (learning rate 0.01, ReLU activation, Adam optimiser) computed on the test set. The confusion matrix indicates the number of observations of the class (true labels indicated in each row) and predicted to be in the class indicated by the columns.}
    \label{tab:conf_mat}
    \small
    \vspace{-5pt}
    \begin{tabular}{l|P{1.2cm}P{1.2cm}P{1.5cm}P{1.5cm}P{1.2cm}P{1.2cm}|ccc}
\toprule
{} &   Up &  Down &  Forward Right &  Forward Left &  Rotate Right &  Rotate Left &  P &  R &  F$_1$ \\
\midrule
Up         &  430 &     3 &      5 &     5 &           5 &          3 & .95 &    .95 &      .95 \\
Down       &    3 &   427 &      4 &     4 &           4 &          3 &       .97 &    .96 &      .96 \\
Forward Right      &    8 &     2 &    544 &     8 &           6 &          0 &       .96 &    .96 &      .96 \\
Forward Left       &    6 &     3 &      7 &   573 &           6 &          0 &       .95 &    .96 &      .96 \\
Rotate Right &    2 &     2 &      2 &     3 &         263 &         55 &       .71 &    .80 &      .75 \\
Rotate Left  &    2 &     3 &      6 &    10 &          86 &        263 &       .81 &    .71 &      .76 \\
\bottomrule
\end{tabular}
\vspace{-10pt}
\end{table}

While the symbol set here only included motion commands, the paper and pen approach allows to easily change the symbols and provides the opportunity to develop more complex commands. While contingent on the creation of a data set with a selection of symbols which must be both intuitive for participants and sufficiently distinct for classification, we do not perceive any technical limitations to expanding the HPL language to include a larger set of instructions (including conditionals and loops for example).

\section{An ER-HPL learning activity}
\label{sec:HPL+ERLS}

\begin{figure}
    \centering
    \vspace{-5pt}
    \includegraphics[width=0.7\linewidth]{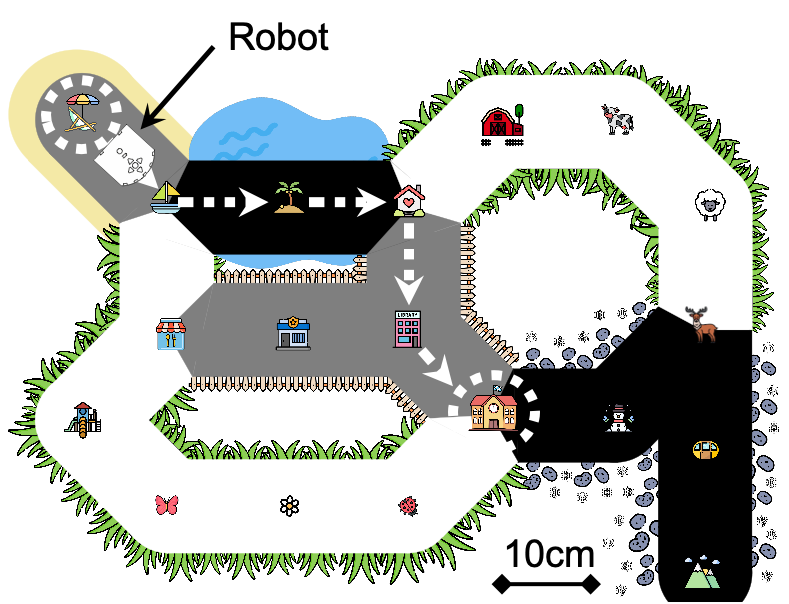}
    \includegraphics[width=0.8\linewidth]{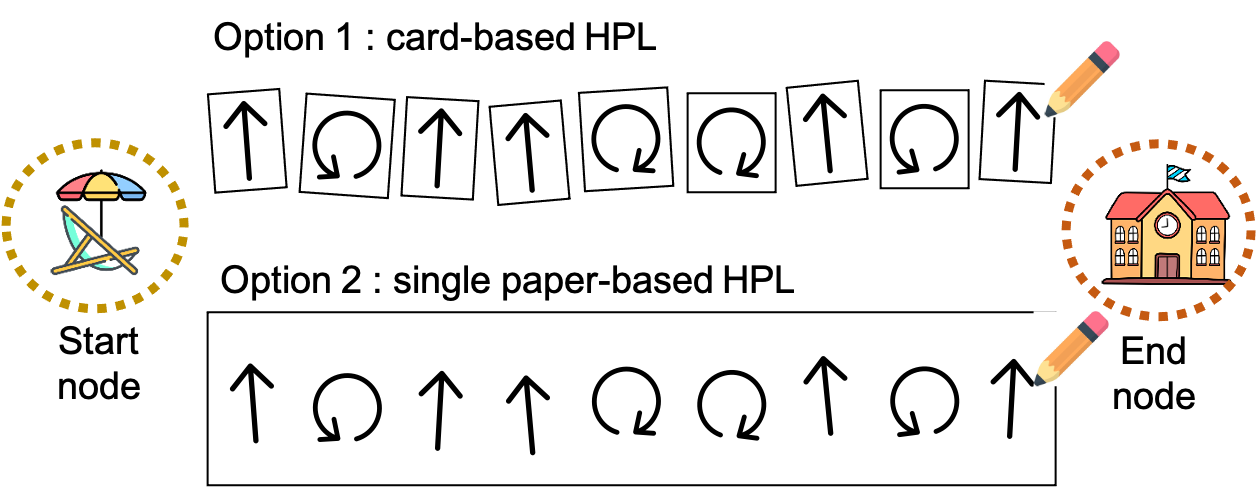}
    \caption{Learning activity with ER-HPL using the Thymio robot. Participants are expected to program the robot to move from one icons on the map to another using the HPL programming language. To learn about the notion of cost, on the playground map (top), the colour of the ground indicates the cost for the robot to navigate along it. The darker the ground colour of a segment, the higher the cost for the robot to navigate it and the faster the robot will use it's initial energy (denoted by the colours of the LEDs on the surface of the robot) using the ground sensors to the intensity of the colour.
    The instructions corresponding to the displayed trajectory (in white on the map) are shown on the bottom. Each step of the algorithm generates either a displacement (indicated by the white arrows on the map) which is equal to the distance between the different illustrated locations on the map (i.e., 11 cm) or a rotation of 45 degrees.
    Icons are taken from FlatIcon and the NounProject.
    }
    \vspace{-10pt}
    \label{fig:HPL_modalities}
\end{figure}

Specifically, the activity aims at helping students 1) understand the concept of sequential programming, 2) discover the notions of cost and path planning, inspired from the work done in the JUSThink project \cite{nasir_robot_2019}, and 3) reinforce the understanding of geometry and angles.
A set of scaffolded tasks were designed to introduce the students to these concepts using the playground and interface presented in Fig. \ref{fig:HPL_modalities}.
In the tasks, students are asked to provide a sequence of instructions allowing the robot to reach a target on a map, by first drawing arrows on sheets of paper and then taking images of them using an ad-hoc application running on a tablet. The application analyses and transforms the instructions into robot motion commands, according to the workflow shown in Fig. \ref{fig:HPL_workflow} - B.

The use of paper introduces the possibility to use a physical exercise book, a natural approach which is close to classroom practices and helps monitor student progress over time. Additionally, it gives the teacher the opportunity to scaffold the instructions according to the needs of the individual students. 
Indeed, to promote collaboration, the teacher can choose to employ a card-based solution (similar to traditional TPLs) requiring students to draw the arrows only once, on separate cards, and then move them around to find the correct sequence (see Fig. \ref{fig:HPL_modalities}, Option 1). Conversely, if the objective is to have the students practice their handwriting, the teacher can choose to have them write the arrows on a single sheet of paper (see Fig. \ref{fig:HPL_modalities}, Option 2).

A preliminary heuristic evaluation of the HPL platform and corresponding activity framework was conducted based on the methodology developed by Giang \cite{giang_towards_2020}. The evaluation included 8 participants with experience in education (2 teachers, 4 engineers, 2 educational researchers) representing the different stakeholders involved in the development of ER content. 
The participants were interested and evoked the engaging, kinaesthetic, playful character of this low cost approach to teaching programming. 
Although one researcher mentioned that they were unsure how the HPL would help improve learning CS concepts, other participants appreciated that the HPL would help train graphomotor skills and increase sense of ownership of the system. 
Participants mentioned that the HPL components were easy to understand (but not necessarily easy to draw), and suggested to continue improving the robustness of the symbol recognition. 
They believed that drawing the symbols on individual cards would facilitate collaboration, and allow to easily modify the number of symbols to be used. 
They believed that the activity set designed was intuitive and didactic. They appreciated the various scaffolding possibilities and the presence of a physical exercise book complementing the in-app instructions. They characterised the overall platform as intuitive, but suggested improvements in terms of storytelling and the application's graphical user interface, specifically recommending additional feedback on the smartphone.

\section{Conclusion}

In this article we propose the \textit{Handwriting-based Programming Language} (HPL), aiming at facilitating the introduction of programming in primary school education. HPL relies on paper and handwriting to program Educational Robots (ER) in a tangible way, while leveraging on well-known practices for early childhood education.  
From the students' perspective, the use of handwriting as input modality introduces the possibility to train graphomotor skills while programming, and can potentially contribute to an increased sense of controllability and ownership, which, like handwriting in general, are important factors for student learning. 
From the teachers' perspective, HPL aims to be close to traditional pedagogical approaches and facilitate the development of learning activities that organically take into account instructions (and instruction modalities), learning artefacts (i.e., the robot, playground and interfaces) and assessment methods \cite{giang_towards_2020}. Therefore, we believe that HPL will be more attractive for teachers who are still reticent about educational technologies. 
Future steps include evaluating these hypotheses in classroom experiments with the target group and expanding to a larger set of instructions. 
One could even envision a fully tablet-based alternative in conjunction with a digital pen and an automated handwriting diagnostic system \cite{asselborn_analysis_2020}. Such a system would not only be able to assess the child's progress over time but would be able to propose adequate remediation to address common handwriting difficulties \cite{asselborn_analysis_2020} through playful means. 

\section{Acknowledgements}

A big thank you goes out to the teacher A.S. and the students involved in the data collection, as well as our always supportive colleagues.

\bibliographystyle{splncs04}
\bibliography{biblio.bib}

\end{document}